# Special Relativity in Absolute Space:
# from a contradiction in terms to an obviousness

Vasco Guerra[†] and Rodrigo de Abreu

*Departamento de Física, Instituto Superior Técnico,
1049-001 Lisboa, Portugal*

**Abstract**

This work deals with the questions of *absolute space* and *relativity*. In particular, an alternative derivation of the effects described by special relativity is provided, which is based on a description that assumes a privileged reference frame. The present theory follows the ideas of Lorentz and Poincaré, abandoning a strict view of Einstein's "equivalence" of all inertial frames. The meaning of the Principle of Relativity is discussed and elucidated, and it is shown that it is *not* incompatible with the existence of a preferred, absolute, frame.

Most scientists nowadays still consider the basic assumptions of the theory proposed here to be plain wrong. Moreover, they tend to see an irreconcilable conflict between the Lorentz-Poincaré and the Einstein-Minkowski formulations. However, as stated by John Bell [Bell1988], although there is a stringent "difference in *philosophy*" between both views, "the facts of physics do not oblige us to accept one philosophy rather than the other". The validity of Bell's assertion is unambiguously demonstrated, and it is shown how and why both approaches do indeed agree in the description of (most of?) the physical phenomena. Evidently, the *physical meaning* of the different physical quantities – such as "time", "speed", "simultaneity" and "synchronization" – is quite different in both programmes. And yet, for perplexing it may look at first sight, the present theory, developed under the Lorentz-Poincaré assumption of a preferred reference frame, somehow encompasses Einstein's theory. There is no conflict, as there is *one* theory.

It must be conceded that what is *said* in both formulations seems to be contradictory, but this is essentially related to a demanding problem of *language*. As a matter of fact, it is revealed that *what special relativity says is not what usually it is thought it says*. By the use of a correct and precise language, problems and paradoxes are immediately avoided. Interpretation problems only arise because words are used in a sense that is often not correct under the chosen description. The core of the problem is related to the largely debated question of synchronization of distant clocks. It is stressed that *reality is not changed by the choices one makes to describe it*, so it is not changed by the particular way in which the clocks have been set.

---

[†] Also at: Centro de Física dos Plasmas, Instituto Superior Técnico, 1049-001 Lisboa

## 1. Introduction

The 20[th] century witnessed a drift in the way of discussing the principles and foundations of physics, towards a more abstract and even technical level. To a big extent, this was a consequence of the development of ever more elaborated physical theories, which require more and more advanced mathematics to their description. Unfortunately, here and then this evolution has gone too far, by focusing the discussion in the mathematical internal consistency of the theories, while forgetting to some extent the underlying physical reality. To "understand" a physical theory has became gradually synonym of "knowing how to perform the calculations", and the theories are often presented to students without a careful discussion of its premises.

However, as noted by J. Resina Rodrigues [Rodrigues1998], the role of mathematics is not to establish the fundamental aspects of reality, but rather to create rigorous formalisms, sets of conclusions derived from *axioms*, valid by themselves as a creation of reason and that can be used as tools by the sciences. On the other hand, physics has no certitude about anything, but has very good *hypotheses*. These hypotheses can be written in terms of elegant mathematics, but do not have the pretension of being the ultimate description of reality. On the contrary, it is assumed the description of nature is always an approximation to the complete truth. Hence, there is a very big difference between the purpose and approach of mathematics and physics. Mathematics makes deductions from a set of axioms, but it does not make sense to ask if these axioms are "valid" in the real world.

This way of thinking has already partly invaded physics. Nonetheless, and quite on the contrary, physics should indeed question and debate its own "axioms", often named principles, and its hypotheses. We have pursued this goal recently, on the subject of special relativity, in our book "Relativity – Einstein's lost frame" [AG2006a]. A big part of the material presented in this article comes from the book. Most calculations are avoided here, and the reader should refer to [AG2006a] for details.

Herein we advocate and demonstrate the compatibility between Einstein's results – based on the notion of *relative* motion – and the existence of a preferred reference frame – with its associated idea of *absolute* motion. The main idea is sketched in [AG2006b]. It is striking the unease revealed by most scientists when discussing the foundations of special relativity and confronted with this statement. One benchmark document to the debate is the work by John Bell [Bell1988]. He has noted that

> Many students never realize, it seems to me, that this primitive attitude, admitting a special system of reference which is experimentally inaccessible, is consistent.
>
> (…)
>
> It is found that if physical laws are Lorentz invariant such moving observers will be unable to detect their motion. As a result it is not possible experimentally to determine which, if either, of two uniformly moving systems, is really at rest, and which is moving.

> The approach of Einstein differs from that of Lorentz in two major ways. There is a difference of philosophy, and a difference of style.
>
> The difference of philosophy is this. Since it is experimentally impossible to say which of two uniformly moving systems is *really* at rest, Einstein declares the notions of 'really resting' and 'really moving' as meaningless. For him only *relative* motion of two or more uniformly moving objects is real. Lorentz, on the other hand, preferred the view that there is indeed a state of *real* rest, defined by the 'aether', even though the laws of physics conspire to prevent us identifying it experimentally. The facts of physics do not oblige us to accept one philosophy rather than the other. And we need not accept Lorentz's philosophy to accept a Lorentzian pedagogy. Its special merit is to drive home lessons that the laws of physics in any *one* reference frame account for all physical phenomena, including the observations of moving observers. And it is often simpler to work in a single frame, rather than to hurry after each moving object in turn.
>
> The difference in style is that instead of inferring the experience of moving observers from known and conjectured laws of physics, Einstein starts from the *hypothesis* that the laws will look the same to all observers in uniform motion.

In this work we adopt not only the Lorentzian pedagogy, but truly the Lorentz's philosophy. We develop Bell's idea and show how and why the facts of physics do not oblige us to accept one philosophy rather than the other. Our assumptions and their subsequent development are close to the ones transmitted by Franco Selleri [Selleri1996, Selleri2005]. Nevertheless, there are evident differences in the presentation, which we have tried to keep extremely simple here. Moreover, the key notion of *Einstein speed* is introduced. Our experience shows that it is extremely difficult to accept and understand the discourse of the present work by keeping always in mind Einstein's philosophy. Because the assertions made in both philosophies may indeed seem to form a *contradiction in terms*. Thus, we urge the readers to advance through this article till the end, with an open mind and no preconceived ideas, while forgetting to a big extent what they already know from Einstein's relativity. This knowledge can be "recovered" with advantage when they reach the end of the text. Hopefully, the compatibility of the present theory with Einstein's special relativity will by then be already considered an *obviousness*. To guide the reader along this conceptual evolution is the purpose of this work.

The structure of this article is as follows. Our starting point is related to Einstein's postulate of the constancy of the *two-way* speed of light in vacuum in all inertial frames. It is further assumed that there is one frame where the *one-way* speed of light in vacuum is the same in all directions of space and equal to $c$, identified with the rest frame, and it is shown this frame is unique. We have denoted this rest frame by *Einstein's frame*.

In section 3 it is established that the one-way speed of light in vacuum is not $c$ in moving inertial frames (the two-way speed of light of course it is) and simultaneity is absolute, contrary to what results in Einstein's relativity. The general expressions for the transformation of coordinates between inertial frames are obtained. They are given by the so-called "synchronized transformation", which differs from the celebrated Lorentz transformation of special relativity.

The meaning of the Principle of Relativity is elucidated and discussed in section 4. In particular, it is shown that the Principle of Relativity is not incompatible with the existence of a preferred, absolute, frame.

As mentioned above, in the present theory simultaneity is absolute, and the same is true for the phenomena of time dilatation and space contraction. The alleged "reciprocity" of these effects made by Einstein's relativity is not "real", and it is explained how and why they may appear to be symmetrical. However, the Lorentz transformation is shown to be *mathematically equivalent* to the synchronized transformation. It is possible and easy to change from "synchronized coordinates" (the description of the phenomena made with the synchronized transformation of coordinates) to "Lorentz coordinates", and vice-versa, thus emerging the compatibility between both theories. We emphasize that, like it or not, this last assertion is a rigorous mathematical statement and as such cannot be questioned.

Clearly, what the Lorentz-Poincaré and Einstein-Minkowski programmes say seems very different. How can they be compatible? After all, they come up with different answers for the questions: is it motion absolute or only relative? are two distant events simultaneous for all observers or not? is the one-way speed of light always $c$ or not? The apparent contradiction is the consequence of a serious problem of *language*. Within Einstein's relativity, the words "time" and "speed", for instance, should be used with a certain sense, perfectly defined by Einstein, but which does not correspond to their intuitive meaning and generally induces an erroneous interpretation of the results. By the use of a correct and precise language, problems and paradoxes are immediately avoided. In particular, the distinction between "speed" and "Einstein speed" is revealed to be crucial.

The core of the problem is related to the old question of synchronization of distant clocks. Usually the subject is discussed in rather abstract terms, and physics is partially lost. Evidently, one can set or "synchronize" his own clocks has it most pleases him, but *reality is not changed by the way the clocks have been set*. Note that this remark should go beyond the standard discussion around the ideas of "conventionalism" and of "operationalism". According to these views, only directly measurable quantities have a physical meaning, the others can *only* be "determined" by human convention. We oppose this view, but a thorough discussion of these matters is left for another opportunity. Additional details can be found in [AG2006a].

Section 5 makes the bridge between the present theory and Einstein's special relativity. The latter is derived from the proposed Lorentzian theory, being clarified what it really says. It is shown that special relativity is incomplete and undetermined unless one actually knows the one-way speed of light or, which is the same, unless the privileged Einstein's frame has been identified.

Section 6 illustrates the compatibility between the present absolute motion theory and Einstein's special relativity through the example of time dilation. Finally, section 7 summarises the main conclusions of this work.

Taking into account our view of special relativity outline above, it was with enthusiasm we read the proposal of Professor Michael Duffy for a book on the "Interpretations of Space-Time Structure". As a matter of fact, our work matched perfectly part of the topics for reflection suggested by Michael Duffy. Furthermore, there is a clear convergence between our proposal and Duffy's own ideas [Duffy2004] in several important topics.

Regarding the original book proposal, our contribution is associated with the Poincaré-Lorentz interpretation of the formal structure of relativity. Following [Duffy2004], our analysis fits two categories of alternative, which are not mutually exclusive: on the one hand, we do state "a preference for the Poincaré-Lorentz formulation as an alternative interpretation of the relativistic formal structure with advantages of its own", exploring an alternative and complete Lorentz programme; on the other hand, we show as well that "the Einstein formulations and the Lorentzian formulations are aspects of the same theory", explaining in detail "how a mathematical formal structure can be given different physical interpretations depending on the concepts used". Notice that one of the striking advantages of the Poincaré-Lorentz formulation presented in this work is that, in spite of its formal mathematical equivalence with Einstein's special relativity, in a sense it *encompasses* the latter theory. This should not be surprising, since a theory of relative movement must be easily described and interpreted, as a special case, in the framework of a theory of absolute motion.

Another remark made in [Duffy2004] is the following: "many colleagues have written papers which are compatible with ether theory, but without employing the word, probably because of the misconceptions surrounding it". This criticism can be made to our presentation, in which we denote the preferred system by Einstein's frame. However, the main reason not to use the word "ether" is that we do not propose any model for the ether itself. Such model requires an even more fundamental approach than the one developed here, and goes outside the scope of the present work. Nonetheless, the possible connection between an ether theory and "quantum vacuum", in the context of quantum field theories, appears to be a relevant direction of research in present day physics.

## 2. Einstein's frame

The study of movement, which is basic to all of physics, has to treat the questions "where?" and "when?". Hence, it is necessary to know how to measure distances and time intervals. Einstein himself explains how to proceed, in the famous 1905 article "On the electrodynamics of moving bodies", where he first presented his Theory of Special Relativity [Einstein1905]:

> The theory to be developed here is based on the kinematics of a rigid body, since the assertions of any such theory have to do with the relations among rigid bodies (coordinate systems) and clocks (...). Consider a coordinate system in which Newton's mechanical equations are valid. To distinguish this system verbally from those to be introduced later, and to make our presentation more precise, we will call it the "rest system."

> If a particle is at rest relative to this coordinate system, its position relative to the latter can be determined by means of rigid measuring rods using the methods of Euclidean geometry and expressed in Cartesian coordinates.
>
> If we want to describe the *motion* of a particle, we give the values of its coordinates as functions of time. However, we must keep in mind that a mathematical description of this kind only has physical meaning if we are already clear as to what we understand here by "time." We have to bear in mind that all our judgments involving time are always judgments about *simultaneous events*.

The scheme is thus very clear: to describe the movement of a certain body, or simply to describe a certain event, it is necessary to give three space coordinates and one time coordinate. Space coordinates are determined by rulers, whereas the time coordinate is determined by clocks. The values of these *coordinates* answer the questions "where?" and "when?". The rest system should then be formed by a grid of coordinates (defined by rigid rods, for example) and a set of "synchronized" clocks in the grid intersection points. We shall not discuss here the question of what rulers and clocks are, nor how do they come into the theory. This matter is raised by Harvey Brown [Brown2005], for instance, and a brief discussion on clocks can be found in [GA2005].

Naturally, the next issue is to know how is it possible to synchronize the clocks of the rest system. This is not too hard and can be done with the help of light signals. It is again Einstein who elucidates how to proceed:

> If there is a clock at point *A* in space, then an observer located at *A* can evaluate the time of events in the immediate vicinity of *A* by finding the positions of the hands of the clock that are simultaneous with these events. If there is another clock at point *B* that in all respects resembles the one at *A*, then the time of events in the immediate vicinity of *B* can be evaluated by an observer at *B*. But it is not possible to compare the time of an event at *A* with one at *B* without a further stipulation. So far we have defined only an "*A*-time" and a "*B*-time", but not a common "time" for *A* and *B*. The latter can now be determined by establishing *by definition* that the "time" required for light to travel from *A* to *B* is equal to the "time" it requires to travel from *B* to *A*. For, suppose a ray of light leaves from *A* to *B* at "*A*-time" $t_A$, is reflected from *B* toward *A* at "*B*-time" $t_B$, and arrives back at *A* at "*A*-time" $t_A'$. The two clocks are synchronous by definition if
>
> $$t_B - t_A = t_A' - t_B \ . \qquad (1)$$
>
> We assume that it is possible for this definition of synchronism to be free of contradictions, and to be so for arbitrarily many points (...).
>
> We have established what is to be understood by synchronous clocks at rest relative to each other and located at different places, and thereby obviously arrived at definitions of "synchronous" and "time". The "time" of an event is the reading obtained simultaneously from a clock at rest that is located at the place of the event, which for all time determinations runs

> synchronously with a specified clock at rest, and indeed with the specified clock.
>
> Based on experience, we further stipulate that the quantity
>
> $$c = \text{(light path)/(time interval)} = 2AB/(t_A' - t_A) \qquad (2)$$
>
> be a universal constant (the velocity of light in empty space).
>
> It is essential that we have defined time by means of clocks at rest in the rest system; because the time just defined is related to the system at rest, we call it "the time of the rest system".

For the sake of clarity of the presentation we have introduced a few minor changes: first, in Einstein's article, the equation does not include the written text, which appears only one page later; second, the equations are not numbered in the original article; finally, the speed of light is denoted by Einstein as $V$ instead of $c$.

Einstein's definition of the rest system is extremely precise and very clear. Taking into account the misinterpretations of Special Relativity that followed Einstein's article and are unfortunately established nowadays, to which Einstein also contributed, it is a bit surprising and very interesting to attest he does *not* define the universal constant $c$ with the *one-way* speed of light, but does so, correctly, with the *two-way* speed of light in vacuum. The latter corresponds to the average speed of light when it makes a round trip, and, contrary to the former (see next section), is indeed a universal constant. In the procedure of synchronizing clocks just described, the time required for light to travel from $A$ to $B$ is related to a one-way speed of light. Of course the time light requires to travel from $B$ to $A$ is also related to a one-way speed of light, but in another direction. It is obvious, but important, that the procedure of synchronization assumes the equality of these two values of the one-way speed of light. From the discussion above, it then follows that

> the rest system, that we shall denote by *Einstein's frame*, is the system in which the one-way speed of light in empty space is $c$ in any direction, independently of the velocity of the source emitting the light.

Notice that since in Einstein's frame the one-way speed of light is known to be $c$, the two clocks $A$ and $B$ can alternatively be synchronized by sending just one light signal from $A$ to $B$, if the distance $L$ between the clocks is known. In fact, if a ray of light leaves from $A$ to $B$ at "$A$-time" $t_A$ and arrives at $B$ at "$B$-time" $t_B$, the two clocks are synchronous if

$$t_B = t_A + L/c. \qquad (3)$$

Although this definition is not as elegant as Einstein's one expressed in (1), it is an equivalent one.

Einstein's frame plays a crucial role in Relativity. It is assumed for now that such frame does exist and further discussion is left for sections 5 and 7. Einstein was fully successful in defining the *rest system* and in establishing a *common time* for it. In the next section we show that if Einstein's frame exists, then it is unique: there is only one frame in which the one-way speed of light in vacuum is $c$ in any direction. That being so, Einstein's frame constitutes a preferred, absolute frame.

### 3. External synchronization and the synchronized transformation

In the previous section we have seen that the questions "where?" and "when?" are answered directly in the rest system with the help of rulers and clocks. More precisely, motion is described by specifying the space and time *coordinates* corresponding to the measurement of space distances and time intervals in Einstein's frame. The next step is to learn how to describe motion, or even just one pair of events, as seen from a different reference frame. It does not look too problematic to do so, and indeed it is not. Notice that the different "objects" that form reality exist independently of the way chosen to describe them. What is necessary is to know the rules allowing the translation from one description to another. Let us find these rules for the case of the description of motion in an inertial frame, *i.e.*, a frame that is moving with a constant velocity in respect to Einstein's frame.

An inertial frame can be devised in quite the same way as Einstein's frame. It is composed as well of a set of rigid rods and synchronized clocks. The rigid rods in both frames are exact copies of each other. This means that when the rods of both frames are brought together, their "meters" have the same size (of course that bringing the rods together requires somehow to stop the moving rulers or to accelerate the ones at rest, or both; as long as the rulers are not deformed when braking – for instance, if they stop by crashing – or accelerating, they always have the same size when brought together, independently of the way used to do so). The same is true for the clocks in both frames. When they are brought together, they have the same rhythm. That being so, any changes in lengths or time intervals that may occur are exclusively induced by the movement.

In order to perform time measurements in the moving inertial frame, it is still required to establish the common time for this frame. In other words, it is necessary to synchronize the moving clocks. This cannot be done in the way used in the rest Einstein's frame, since in the moving frame the one-way speed of light is not known[1]. Nevertheless, it can easily be done with the help of the clocks at rest, because these clocks have already been synchronized. Hence, the moving clocks can be synchronized simply by adjusting them to zero whenever they fly past a clock at rest that shows zero as well. From that moment on the moving clocks remain synchronous between themselves, thus establishing the common time of the moving system. Evidently, this synchronization procedure is not the standard one. We shall not start here any discussion around the "conventionality of synchronization", which has ample literature available and is only briefly addressed bellow.

Let $S$ denote the rest system and $S'$ the moving one. For simplicity, assume the axis of both frames are aligned, and that the origin of $S'$ moves along the *x*-axis of $S$ with a certain speed $v$, in the positive direction. The primed and non-primed quantities correspond to measurements made with the rulers and clocks of $S'$ and $S$, respectively. The synchronization procedure just delineated corresponds simply to the statement that $t=0$ implies $t'=0$, as it is shown in figure 1. Of course this synchronization method is an

---

[1] This sentence may be a shocking one. Of course that in Einstein's relativity "synchronization" is done *as if* the one-way speed of light is $c$, i.e., "stipulating" its value to be $c$. The issue is discussed in section 5.

*external* one, as noted by Mansouri and Sexl in [MS1977], since to synchronize the clocks from *S'* one has to use the clocks from *S*.

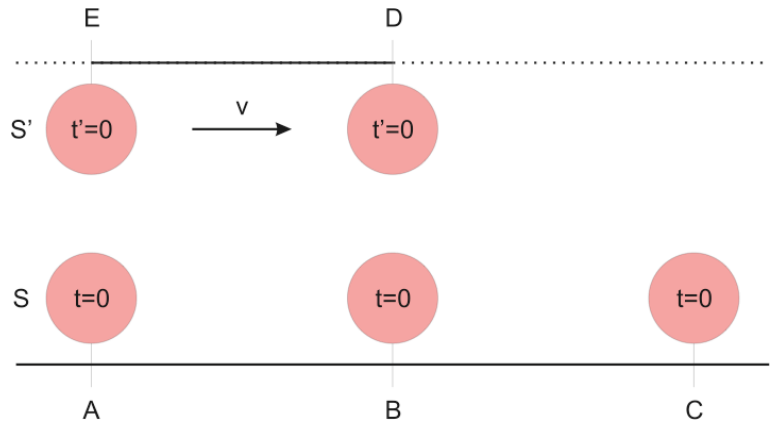

Figure 1: Synchronization of clocks in a moving system: the moving clocks *D* and *E* are synchronized with the help of the previously synchronized clocks at rest *A* and *B*.

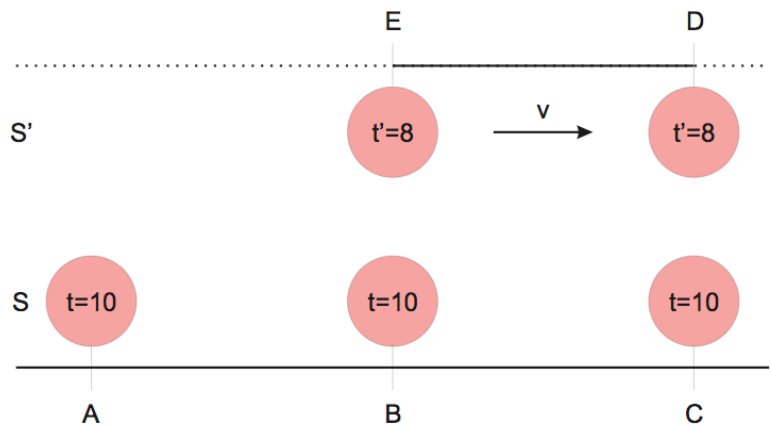

Figure 2: Evolution of the situation from figure 1 after 10 milliseconds. All times in the figure are expressed in ms.

With rulers and synchronized clocks equipping the moving frame, it is now possible to study motion from it. Distances and time intervals can be measured. However, it is of course desirable to be able to relate them to the distances and time intervals measured in Einstein's frame.

The phenomenon of time dilation can be deduced in the usual way, such as presented in the classic textbooks from Feynman [FLS1979] or Serway [SB2000], using a light clock placed in *S'* aligned along the *y*-axis. The well-known result

$$\Delta t = \gamma \, \Delta t' \qquad (4)$$

with

$$\gamma = 1/(1 - v^2/c^2)^{1/2} \qquad (5)$$

expresses the fact that "moving clocks run slower". Take note that the phenomenon is independent of synchronization, as *each* of the moving clocks experiences time dilation. Referring to figure 1, if $v=0.6c$ (so that $\gamma=1.25$) and if the distances between clocks *A* and *B* and between clocks *B* and *C* are the same, *L*, and equal to 1800 km, the situation depicted in figure 1 evolves to the one depicted in figure 2 at t=10 ms. Time dilatation is evident from the figures. While in the rest system 10 ms have passed, in the moving frame only 8 ms passed. But the way to correlate both descriptions is known. Someone in the moving frame may say "clock *D* took 8 ms to go from clock *B* to clock *C*". An observer at rest will agree on the reality that is being described, the movement of clock *D* from *B* to *C*. And will know that for him clock *D* took 10 ms to do this trip, even without measuring himself the time duration of the trip. Notice that the effect is induced by absolute motion and clearly there is no reciprocity of time dilation (!).

The phenomenon of space contraction can be deduced as well in the usual way, using a light clock placed in *S'* aligned along the *x*-axis. The well-known result

$$L' = \gamma L \qquad (6)$$

expresses the fact that "moving rulers are shorter" (the observers in *S'* measure a bigger length, simply because their "meters" have become shorter). Referring to figures 1 and 2, the distance in *S'* between clocks *D* and *E* is $L'=\gamma L=1.25 \times 1800$ km = 2250 km. Again, the effect is induced by absolute motion and clearly there is no reciprocity of space contraction.

The transformation of coordinates between *S* and *S'* can now be readily obtained. If the origins of both frames are considered to be at clocks *A* and *E*, the position *x'* of clock *D* in *S'* is simply given by

$$x' = L' = \gamma L = \gamma (x - vt) \qquad (7)$$

being *x* its position in *S*. Consequently, the relations between space and time coordinates providing the translation from the description in the rest system to the one in a moving frame are just given by

$$x' = \gamma (x - vt)$$
$$t' = t/\gamma \qquad (8)$$

where $\gamma$ is given by equation (5) and *v* is the absolute speed (*i.e.*, the speed measured in the rest system) of the moving frame.

Expressions (8) form the *synchronized transformation*. They were obtained by Mansouri and Sexl in 1977 [MS1977] and by Abreu and Homem in 2002 [Abreu2002, Homem2003], and have been emphasized by Franco Selleri since 1996 [Selleri1996, Selleri2005], who named them as *inertial transformations*. Interestingly enough, the synchronized transformation is not symmetrical, as the inverse transformation, expressing *x'* and *t'* as functions of *x* and *t*, is given by,

$$x = 1/\gamma (x' + \gamma^2 vt')$$
$$t = \gamma t' \qquad (9)$$

Notice that the position of the origin of S, x=0, is given in S' by $x'=-\gamma^2 vt'$. This means that S' sees S passing with speed $v'=-\gamma^2 v$, and not just -v as one could think at first sight. One factor γ accounts for the fact that rulers are shorter in S', while the second γ factor comes from the fact that clocks run slower there.

It is not difficult to derive that if an object goes with absolute speed w, then its relative speed, $w_v$, in relation to a frame S' moving with absolute speed v, is given by

$$w_v = \gamma^2 (w - v) = (w - v) / (1 - v^2/c^2) \qquad (10)$$

Thus, it is possible to calculate the one-way speed of light in a moving frame. If a light ray is emitted and travels in the positive direction of the x-axis, we know it propagates in the rest system with speed c, independently of the speed of the source emitting the ray. In S', the one-way speed of this light ray is given by (10) with w=c,

$$c_v^+ = \gamma^2 (c - v) \qquad (11)$$

If the light ray is emitted in the negative direction of the x-axis, w=-c and its speed is given, in absolute value, by

$$c_v^- = \gamma^2 (c + v) \qquad (12)$$

Therefore, following the Lorentz's philosophy as developed here, in the moving frame the one-way speed of light is not the same in different directions and $c_v^-$ is always bigger than c. And yet the two-way speed of light is always c, as the reader can easily verify. The detailed calculations leading to expressions (4 – 12) can be found in [AG2006a], as well as further developments of the proposed approach (including the transformation of coordinates between two moving inertial frames and additional insight on the velocity addition expressions).

Besides absolute time dilation and absolute space contraction, the synchronized transformation exhibits as well absolute simultaneity. How can these statements be compatible with Special Relativity, which claims the relativity of all these effects? The answer is given in the next two sections.

## 4. The principle of relativity

The principle of relativity is so important to physics that it gave its name to Einstein's theory of relativity. It is often lightly associated with the idea that "nothing is absolute and everything is relative", including motion, since things depend from the observer's point of view. For instance, a passenger sitting in a moving train is at rest in relation to the train, but in motion in relation to the ground. Therefore, it is argued, it is not possible to speak about absolute motion nor absolute rest and all inertial frames are "equivalent". Feynman made a hilarious but interesting observation [FLS1979]:

> The fact that "things depend upon your frame of reference" is supposed to have had a profound effect on modern thought. One might well wonder why, because, after all, that things depend upon one's point of view is so simple an idea that it certainly cannot have been necessary to go to all the trouble of the physical relativity theory to discover it. That what one sees depends upon his frame of reference is certainly known to anybody who

> walks around, because he sees an approaching pedestrian first from the front and then from the back; there is nothing deeper in most of the philosophy which is said to have come from the theory of relativity than the remark that "A person looks different from the front than from the back".

Obviously the principle of relativity goes much deeper than Feynman's joke and goes much deeper than "motion is relative". The point is to know if motion is *only* a relative notion (Einstein's philosophy) or if it can be given an *absolute* meaning as well (Lorentz's philosophy, see Bell's quote in section 1). The fact that one can speak about relative motion, like the passenger that is at rest in relation to the train, does not imply immediately that one cannot speak about absolute motion. Similarly, the fact that one can describe reality from different inertial frames in a similar way does not imply they are all "equivalent".

Let us try to debate what does the principle of relativity really mean. It seems interesting to start with a brief presentation of its historical evolution. The goal is to show how and why the principle of relativity *is* indeed consistent with the ideas developed in the previous section, associated with the existence of a preferred reference frame. Moreover, the next section also clarifies how these ideas relate to Einstein's theory of special relativity.

The origins of the principle of relativity are usually attributed to Galileo and to his 1632 "Dialogue Concerning the Two Chief World Systems" [Galilei1632], although, as pointed out by Roberto A. Martins [Martins1986], very similar arguments have been previously used by Nicole Oresme in 1377, Giordano Bruno in 1584 and by Galileo himself in 1624 (in a letter to priest Francesco Ignoli). Anyway, in the defence of the heliocentric system, Galileo argued it is not possible to conduct a physical experiment capable of indicating if a body is immobile or in motion. He used in the "Dialogue" a famous metaphor with a ship, Sagredo concluding:

> I am therefore satisfied that no experiment that can be done in a closed cabin can determine the speed or direction of motion of a ship in steady motion.

Galileo mentioned the need to be "below decks" (not quoted here) and "in a closed cabin". He used this example in order to show that one cannot determine whether the earth is revolving or fixed, in the same way that from the motion of butterflies one cannot determine if a ship is moving or standing still. It is often considered he wanted to stress the idea that there is no meaning in the concept of a moving body without reference to its movement *relative* to another body. However, if it would be only a question of relative movement, what difference would it make to consider either the earth moving or the sun moving? Galileo felt the difficulty in arguing only with relative motion. He was (correctly) convinced that saying the earth revolves and the sun is immobile is closer to reality than the reverse, but the truth is that he could *not* find a strong and unquestionable argument to support his view. His purpose was "only" to show that the everyday experience was not in contradiction *neither* with an earth moving *nor* with an immobile earth. That the effects observed in experiments performed on earth give the same result regardless of what motion the earth really has, being thus impossible, by experiments performed on earth, to solve the doubt[2]. In fact, Galileo was going after the idea of *absolute* motion and *absolute* rest (!). He argued it is difficult to decide if it is the earth or the sun that is at *absolute* rest, because we can only perceive relative motion. Even though, after the work of Galileo it has been generally accepted that the notions of "rest" and "movement" are strictly *relative*, velocity having meaning *only* as "relative velocity".

This is something very different from the notions developed in the previous sections, where it was argued velocity – and therefore motion – can be given an *absolute* meaning. Galileo has shown the importance of relative motion, but there is no inconsistency between the notion of absolute rest and Galileo's work. Quite on the contrary! So it is not at all accurate to mention Galileo and then jump into conclusions against the concept of absolute motion. Hence, it is necessary to proceed with care and avoid the temptation of naive or void assertions. Galileo has most certainly seen beyond the simplistic statement "motion is only a relative notion". His *principle of relativity* is summarized in the quote above: just with an experiment conducted inside a closed cabin it is not possible to decide if the ship is at rest or in steady motion. But there is an important observation, which is the mentioned comment about the need to be below decks. In this chapter it will be seen how the principle of relativity, related to the impossibility of detecting the ship's motion inside the cabin, *is compatible* with the reality of a preferred, and thus absolute, frame. And we can advance the key point is connected precisely with the need to be "below decks".

The principle of relativity was first enunciated by Newton in 1729 [Newton1729], in the line of Galileo's example:

> the motions of bodies included in a given space are the same among themselves, whether that space is at rest, or moves uniformly forwards in a right line without any circular motion.

Contrary to Galileo, who argued that any "common motion is as non-existing" (and this included earth's rotation!), Newton makes the important restriction of considering only uniform motion along straight lines, *i.e.*, inertial frames. To Newton there existed one absolute space and a multitude of inertial systems. Even if empirical observations cannot detect if a certain body is at absolute rest, Newton felt the need to introduce this concept anyway. He needed the concepts of absolute space and absolute rest in order to state the first law of motion: that every body continues in its state of rest, or of uniform motion in a right line, unless it is compelled to change that state by forces impressed upon it. Without the assumption of absolute space no meaning can be given to the notion of absolute rest, which seemed to Newton, as it did somehow to Galileo, a fundamental experience that could not be dispensed within the formulation of the first law of motion. Therefore, Isaac Newton founded classical mechanics on the view that space is something distinct from the bodies. He distinguished the basic notion of space from the various ways by which we measure it; the former he called absolute space and the latter relative spaces. He subsequently defined the true motion of a body to be its motion through this absolute space, which is thus the "stage of reality".

---

[2] It is worth noting that Galileo's argumentation about the two chief world systems, Ptolemaic & Copernican, is not very far from our the thesis of this text, about the Lorentz-Poincaré and Einstein-Minkowski approaches to special relativity!

Einstein's theory of relativity contributed decisively to the progressive abandon of the notion of absolute space in favour of that of "equivalence" between all inertial frames. That being so, it is particularly striking to see that Einstein did not accept a straightforward denial of absolute space. As a matter of fact in 1953, two years before Einstein's death, he wrote [Jammer1994]:

> two concepts of space may be contrasted as follows: (a) space as positional quality of the material objects; (b) space as container of all material objects. In case (a), space without material object is inconceivable, in case (b), a material object can only be conceived as existing in space; space then appears as a reality which is in a certain sense superior to the material world. The concept of space was enriched and complicated by Galileo and Newton, in that space must be introduced as the independent cause of the inertial behaviour of bodies if one wishes to give the classical principles of inertia (and herewith the classical law of motion) an exact meaning. To have realized this fully and clearly is in my opinion one of Newton's greatest achievements.

Of course concept (b) is no more no less than Newton's stage of reality. Einstein argued in favour of concept (a) and nowadays the idea of absolute space is essentially considered as erroneous, or at best superfluous. Nevertheless, for one reason or another, neither Galileo, Newton, nor even Einstein have completely ruled it out.

Historically, it was of major importance that the laws of classical mechanics look identical in all moving inertial frames and in the rest system under Galileo's transformation of coordinates. For example, in some inertial frame $S'$ Newton's second law for the $x$-component reads

$$F_x' = m \, d^2x'/dt'^2 \qquad (13)$$

where $F_x'$, $m$ and $d^2x'/dt'^2$ denote the component of the force along $x'$, mass and the component of acceleration along $x'$ in $S'$, respectively. It is irrelevant to discuss this law here. What is important to emphasize is that if space and time coordinates are transformed to a second inertial frame $S''$ according to the Galileo transformation, then Newton's second law *keeps exactly the same form* in $S''$,

$$F_x'' = m \, d^2x''/dt''^2 \qquad (14)$$

where $F_x''$, $m$ and $d^2x''/dt''^2$ denote the component of the force along $x''$, mass and the component of acceleration along $x''$ in $S''$, respectively. Poincaré [Poincare1904] and Einstein [Einstein1905] generalized this idea to *all* laws of physics. Poincaré includes the principle of relativity among the "five or six general principles to the various physical phenomena",

> The laws of physical phenomena must be the same for a 'fixed' observer
> as for an observer who has a uniform motion of translation relative to him,
> so that we have not, and cannot possibly have, any means of discerning
> whether we are, or are not, carried along by such a motion.

Very remarkably, Poincaré's principle of relativity is formulated under the assumption of absolute space. To Poincaré, that one cannot have means to detect absolute motion is not contradictory with its existence. All frames *appear* to be equivalent, even if they are not. Feynman called it a *nature conspiracy* [FLS1979]: since any experiments devised to measure an absolute speed *u* seem to have failed,

> it appeared nature was in a "conspiracy" to thwart man by introducing some new phenomenon to undo every phenomenon that he thought would permit a measurement of *u*.

> It was ultimately recognized, as Poincaré pointed out, that *a complete conspiracy is itself a law of nature!* Poincaré then proposed that there *is* such law of nature (...); that is, there is no way to determine an absolute speed.

Soon after Poincaré, Einstein formulates the principle of relativity in the form

> not only the phenomena of mechanics but also those of electrodynamics have no properties that correspond to the concept of absolute rest. Rather, the same laws of electrodynamics and optics will be valid for all coordinate systems in which the equations of mechanics hold.

After Galileo's epic struggle in favour of the heliocentric system and 100 years of Einstein's celebrated theory of relativity, many textbooks on elementary physics state the principle of relativity more or less on Einstein terms, by stressing first that the laws of physics must be the same in all inertial reference frames and that all inertial frames are "equivalent". From here it is secondly concluded the principle of relativity asserts that there is no physical way to distinguish between a body moving at a constant speed and an immobile body: it is of course possible to determine that one body is moving relative to the other, but it is impossible to determine which of them is moving and which is immobile.

There is a subtle but critical difference in both assertions, as the fact that the laws of physics "keep the same form" in all frames has nothing to do with the principle of relativity. One of the misconceptions with the principle of relativity is this confusion between both statements. In its genesis the relativity principle is a "principle of relative movement", hence its name. It is solely related to the impossibility of detecting absolute motion. But very often, it is believed it truly corresponds to saying the laws of physics keep the same form in all inertial frames and to the equivalence of all these frames. One noticeable exception among physics textbooks is "Feynman lectures on physics" [FLS1979]. Feynman recovers the idea of a "principle of relative motion" and puts it luminously, with an important and barely seen observation, noted here with italics:

> if a space ship is drifting along at a uniform speed, all experiments performed in the space ship will appear the same as if the ship were not moving, *provided, of course, that one does not look outside*. This is the meaning of the principle of relativity.

That one cannot look outside is the equivalent of Galileo's remark about being below decks. Here we will take the principle of relativity as enunciated by Feynman, with the additional constraint of being in vacuum (discussed in depth in [AG2006a]):

> All the experiments performed in a closed cabin in vacuum in any moving inertial frame will appear the same as if performed in Einstein's frame, provided, of course, that one does not look outside.

Notice that the interdiction of looking outside raises an extremely delicate point. In fact, in the very construction of a moving inertial frame as presented in section 3, we *have to look outside* in order to synchronize the moving clocks by comparison with the clocks of Einstein's frame. In this sense, the procedure of synchronization is an *external* one. This fact casts a new light into the meaning of the principle of relativity, which we will develop in the remaining of this work. For the moment, let us simply note that [FLS1979]

> Our inability to detect absolute motion is a result of *experiment* and not a result of plain thought (...). There is a philosophy which says that one cannot detect *any* motion except by looking outside. It is simply not true in physics. True, one cannot perceive a *uniform* motion in a *straight line*, but if the whole room were *rotating* we would certainly know it (...). Therefore it is not true that "all is relative"; it is only *uniform velocity* that cannot be detected without looking outside. Uniform *rotation* about a fixed axis *can* be.

This is why Galileo was right in saying the earth rotates and not the sun: because rotation can be given an *absolute* meaning[3]. We state that uniform motion can be given an absolute meaning as well, even if the principle of relativity (in vacuum) suggests it is impossible to identify Einstein's frame. However, as it is indicated in section 7, it seems it may be indeed possible to "look outside" and to determine which is Einstein's frame. In any case, the crucial idea is that the principle of relativity does not say "all is relative" and it is by no means incompatible with the notion of absolute space.

The remaining of this section may seem at first a deviation on the road leading to a deeper understanding of the principle of relativity. However, it is absolutely necessary and introduces the Lorentz transformation and the key notion of "Einstein speed".

---

In [AG2006a, AG2006b] it is shown how to obtain a formal Galileo transformation of coordinates between inertial frames, which is mathematically equivalent to the synchronized transformation (8) and, hence, can be used to study motion in inertial frames. The interested reader can see those references for details. Here we shall focus only on the Lorentz transformation, which provides yet another way to relate the space and time coordinates of the rest system $S$ to the ones of a moving inertial frame $S'$.

Once the clocks in both frames have been synchronized as described in section 2, the Lorentz transformation can be easily obtained by "correcting" in a particular way the time readings of the moving clocks. Notice again that the synchronization scheme is *external*, since it requires the observers from $S'$ to look at $S$.

---

[3] The sun actually *also* rotates, but it rotates less than the earth does. Anyway, the motion of the sun along a curved path has nothing to do with the circular motion it exhibits on earth's sky.

In [AG2006a, AG2006b], the Galileo transformation was attained by defining shorter seconds for the moving clocks. Now the rhythm of the clocks will not be changed, only they will start from a different condition. Instead of adjusting the moving clocks to mark $t'=0$ when $t=0$ (see figure 1), we shall do that to one clock of $S'$ only, which identifies the position $x'=0$. The remaining moving clocks will be delayed by a factor that is proportional to their distance $x'$ to the reference position $x'=0$, which is given by $(v/c^2)x'$ (if $x'$ is negative, this corresponds actually to advancing the clock). We shall denote the clocks altered in this way by *Lorentzian clocks*, and their time readings, $t_L'$, by Lorentzian times. We thus have

$$t_L' = t' - (v/c^2)x' \qquad (15)$$

Why should someone be interested in "de-synchronizing" clocks according to (15) will be clarified in the next section and it is related to the problem of performing an *internal* "synchronization" of the moving clocks.

Referring to figure 1, suppose clock $E$ defines the position $x'=0$. Then, clock $E$ marks $t_L'=0$ when $t=0$. Since clock $D$ is located at $x'=2250$ km and $v=0.6c$ (*cf.* section 2), it must be delayed $(v/c^2)x' = (0.6/c) \times 2250 \times 10^3 \approx 0.0045$ s = 4.5 ms. Therefore, at $t=0$ clock $D$ reads $t_L'=-4.5$ ms. Similarly, a clock $F$ (not shown in the figure) placed at the same distance from $E$ to the left, at $x'=-2250$ km, would show $t_L'=+4.5$ ms when $t=0$. Notice that the moving clocks $D$ and $E$ are exactly equal to clocks at rest $A$, $B$ and $C$, only they are not synchronized as to mark all $t_L'=0$ at some arbitrary instant. The situation is shown in the upper part of figure 3. Since the moving clocks are precisely the same as in figures 1 and 2, they exhibit strictly the same time dilation as shown in those figures. Their *rhythms* are affected by time dilation exactly as before. The only difference is that now their starting condition was set in a different way. Hence, when 10 ms have passed in $S$, only 8 ms elapsed in $S'$. More precisely, at $t=10$ ms clocks $D$ and $E$ mark $t_L'=-4.5+8=3.5$ ms and $t_L'=0+8=8$ ms, respectively, as represented in the lower part of figure 3. Clock $F$, not shown and now on top of clock $A$, would mark $t_L'=4.5+8=12.5$ ms. Notice that *figure 3 displays exactly the same reality as figures 1 and 2*, which is simply *described* in a different way. The positions of $D$ and $E$ can have two clocks each, one synchronized clock as in figures 1 and 2, and one Lorentzian clock as in figure 3. Therefore, one description does *not* oppose the other: they both coexist and they both can be used.

With Lorentzian times, the expressions for transformation of coordinates between Einstein's frame and the moving frame are easily found by substituting $t'$ and $x'$ given by the synchronized transformation (8) into (15):

$$t_L' = t/\gamma - (v/c^2)\gamma(x - vt) = \gamma[t/\gamma^2 - (v/c^2)x + (v^2/c^2)t] \qquad (16)$$

Substituting $\gamma$, the *Lorentz transformation* is finally obtained,

$$x' = \gamma(x - vt)$$
$$t_L' = \gamma[t - (v/c^2)x] \qquad (17)$$

with $\gamma$ given by (5) and $v$ denoting the absolute speed of $S'$.

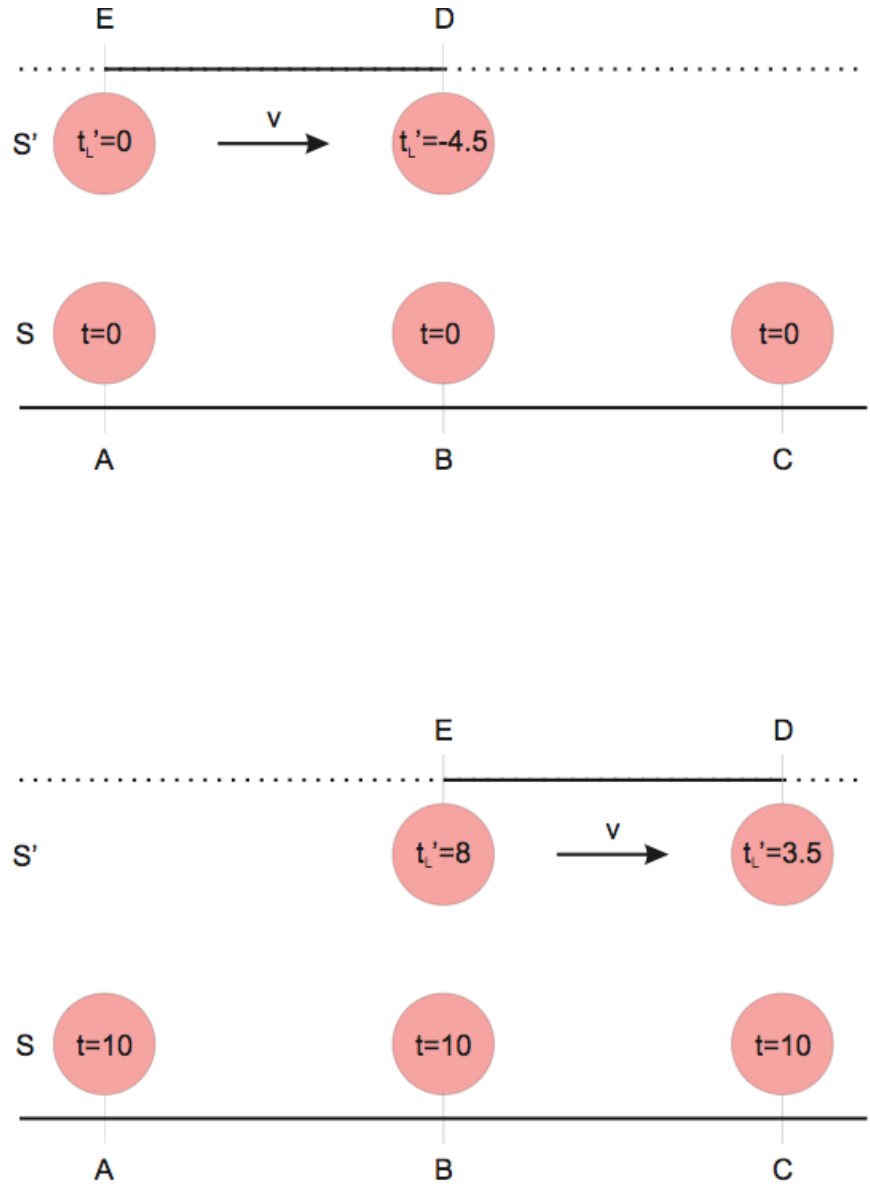

Figure 3: Time dilatation with de-synchronized and equal clocks. All times expressed in milliseconds.

Clearly, it is possible to *define* an *Einstein speed*, $v_E$, as the "speed" measured with Lorentzian clocks (and ordinary rulers),

$$v_E = \Delta x' / \Delta t_L \qquad (18)$$

The time interval is calculated as the difference of the "time reading of a clock located at arrival position" with the "time reading of a clock located at departure position". Since Lorentzian clocks are de-synchronized, Einstein speeds are of course different from speeds (which can be calculated by $v = \Delta x'/\Delta t'$ when the time intervals and distances are measured with synchronized clocks). Nevertheless, two-ways speeds of any object are the same with both types of clocks, since they are measured with one clock only and, that being so, any de-synchronization of distant clocks has no effect in the measurements.

It is easy to find the expressions relating Einstein relative speeds with the true relative speeds. Substituting relation (15) into

$$x' = w_v t' \qquad (19)$$

where $w_v$ is the (relative) speed of an object whose absolute speed is $w$, measured in frame $S'$ that goes with absolute speed $v$, and comparing with (18), the Einstein velocity $v_E$, measured in a frame moving with absolute speed $v$, of an object which has absolute speed $w$, is

$$v_E = w_v/(1 - vw_v/c^2) = (w - v)/(1 - vw/c^2) \qquad (20)$$

The Einstein speed of light, $c_E$, exhibits a very interesting property. As a matter of fact, since the absolute speed of light is always $c$, $c_E$ is obtained directly from (20) with $w=c$:

$$c_E = (c - v)/(1 - v/c) = c \qquad (21)$$

Therefore, *the Einstein speed of light is always c in any moving inertial frame, independently of the speed of the moving frame*. This result is simple, but very important. The distinction between speed and "Einstein speed" must be made. The one-way speed of light is *not c* in all inertial frames [*cf.* (11) and (12)]; however, the one-way Einstein speed of light is *c* in all inertial frames. This will be further analysed in the next section.

Let us stress once more that figure 3 and figures 1 and 2 show precisely the *same reality*, which is simply *described* in a different way. Nothing prevents the observers in $S'$ from describing all events using Lorentzian (or other) clocks. A few problems may arise, though. According to the procedure of clock synchronization devised in section 2, Lorentzian clocks are de-synchronized. Therefore, two clocks placed in distinct locations working well in a certain frame cannot be "transferred" to another frame and be expected to work well. They will mark wrong Lorentzian times. Before they can be used they must be corrected with the appropriate de-synchronization factors. Only then they can be utilized as Lorentzian clocks. For instance, if clocks $A$ and $B$ from figure 3 are given to the moving frame $S'$ at $t=0$ (to locations $E$ and $D$) they will mark wrong Lorentzian times. A second important question is again related to the *meaning* of the quantities measured with Lorentzian clocks. If Lorentzian times are taken as "times", the most obvious of these problems is with the notion of "simultaneity", which determines the common time of a certain frame. Suppose that two events, occurring in different space locations, are *defined* as "simultaneous" in a certain frame if the time coordinates given by the clocks of that frame, placed in the locations where the events take place, are the same. In that case, two events taking place at points $A$ and $B$ and $t=0$ are "simultaneous" in $S$, since clocks $A$ and $B$ show the same time readings ($t=0$). Are the same events considered to be simultaneous in $S'$ as well? If the moving clocks are truly synchronized, *i.e.*, if they have been set according to the external synchronization procedure from section 2 and corresponding to figure 1, the answer is "yes", since clocks $D$ and $E$ both mark $t'=0$. But if the moving clocks are Lorentzian clocks, then clocks $D$ and $E$ show different *numbers*, repectively $t_L'=-4.5$ and $t_L'=0$ ms (see figure 3). If these numbers, which are just the time *coordinates* used to describe events, are interpreted as "true" times, then the two events are *not* taken as simultaneous in the moving frame *S*", so that the answer is "no". Clearly the difficulties lay in the interpretation that is made of Lorentzian times and in a *question of language*. As a matter of fact, the word "simultaneous", although precisely defined, was used with two different meanings. There are actually *infinite* different *definitions* of "simultaneity", corresponding to distinct ways of de-synchronizing the moving clocks in variants of expression (15).

Interestingly enough, a rather similar point has been raised by Edwards back in 1963 [Edwards1963]:

> There are an infinite number of correct transformations corresponding with the infinity of possible clock synchronizations. (...) The only *real* difference in the forms of these transformations is in how one wishes to synchronize clocks. (...) To mention the fact that two spatially non-copunctual events simultaneous in one system are not simultaneous in a system moving with constant relative velocity is now standard in textbooks. Now, however, it is easy to show that with a proper setting of clocks in the *S*' system, two events simultaneous in one system can *always* be made simultaneous in a second regardless of the value of the relative velocity.

It may be somewhat surprising that the Lorentz transformation and other de-synchronized transformations, with their different definitions of simultaneity, are *mathematically equivalent* to the synchronized transformation (8). If we know the Lorentzian coordinates of a certain event, then we can immediately know its synchronized coordinates, and vice-versa, as long as the absolute speed of the moving frame is known. Notice that for the point we are trying to make here, it is irrelevant if this speed is experimentally inaccessible or not (see also section 5).

Two peculiarities of the Lorentz transformation are still very interesting to note. The first one is that, contrary to the synchronized transformation, the Lorentz transformation is *symmetrical*. As a matter of fact, the quantities in Einstein's frame can be expressed as a function of the Lorentzian ones by inverting (17), resulting

$$x = \gamma (x' + vt_L')$$
$$t = \gamma [t_L' + (v/c^2)x] \qquad (22)$$

with $\gamma$ given by (5), rewritten here,

$$\gamma = 1/(1 - v^2/c^2)^{1/2} \qquad (5)$$

and *v* denoting the absolute speed of *S*'. This set of equations is the same as (17), simply interchanging the roles of the quantities in both frames and replacing *v* by -*v*. The position of the origin of *S*, *x*=0, is given in *S*' by $x'=-v\, t_L'$, so that *S*' sees *S* passing with *Einstein speed* -*v*. In what concerns Einstein speeds, there is no difference in the ways "*S* sees *S*'" and "*S*' sees *S*". The second observation is that if a second inertial frame *S*" goes with absolute speed *w*, then the transformation of Lorentzian coordinates between *S*' and *S*" *takes the same form* as between Einstein's frame and a moving inertial frame,

$$x' = \gamma_E (x'' + v_E t_L'')$$

$$t_L' = \gamma_E [t_L'' + (v_E/c^2)x''] \qquad (23)$$

with $\gamma_E$ given by

$$\gamma_E = 1/(1 - v_E^2/c^2)^{1/2} \qquad (24)$$

and

$v_E = (w-v)/(1-vw/c^2)$ being the *relative* Einstein velocity between both moving frames. These two facts could eventually suggest all moving inertial frames are equivalent to Einstein's frame and only relative motion is of importance. If that would be the case, Einstein's frame would not be a privileged frame after all. But this "equivalence" is purely formal. It emerges as a consequence of using Lorentzian clocks. A similar "equivalence" emerges from the use of Galilean clocks as defined in [AG2006a, AG2006b], which shows that it is not necessary to use the Lorentz transformation to achieve this kind of formal "equivalence" of all inertial frames.

A deeper understanding of the principle of relativity is obtained from Einstein's special relativity, as it is discussed in the next section.

## 5. Einstein's special relativity

In section 2 it has been seen how to synchronize clocks in Einstein's frame with the help of light signals. The synchronization of clocks in a moving frame was subsequently detailed, the procedure involving "looking outside" from the moving frame to Einstein's frame. This synchronization scheme was therefore named as *external* synchronization and is associated to the intuitive ideas of synchronization and simultaneity. Section 4, however, deals with the principle of relativity, which asserts it is impossible to detect absolute motion (in vacuum) without looking outside. The principle of relativity raises hence the question of what can be done to somehow "synchronize" moving clocks without looking outside, *i.e.*, to perform some *internal* "synchronization". Of course "synchronization" then becomes a dangerous word to use and has the same problem pointed out with the word "simultaneous" in the last section. What is necessary is to find a way to give some well-defined starting condition for all the clocks in a particular frame. Those conditions can be *defined* as clock "synchronization", although they may not correspond to the everyday notion of a true synchronization. For instance, the de-synchronized Lorentzian clocks can be defined as being synchronized. It may be confusing, but there is no problem at all in doing that. Lorentzian clocks are good enough to make time measurements. As long as one knows, of course, what kind of clocks is being used. Here, the word synchronization will be used as synonym of external synchronization, other synchronization definitions being always precisely identified. For a while they will be emphasized by the use of " " signs.

In the rest system *S*, the stage of reality coinciding with Einstein's frame, there is no need to look outside. Since the one-way speed of light is known to be *c* in every direction, the synchronization procedure of the clocks at rest is in fact an internal one. Now, suppose an inertial frame *S'* is moving with absolute speed *v*. If the observers in *S'* cannot look outside, they do not know they are moving. How can they "synchronize" internally their clocks? Well, they can just carry on *as if* they were at rest! They can simply *assume* the one-way light speed to be *c* in every direction in their own frame, although it is not, and then make the internal "synchronization" of their clocks consistent with this assumption.

Suppose a light ray is emitted from *S'* at a certain point *E* and at $t_E'=0$, and travels a distance *L'* until it reaches a second point *D*. To internally "synchronize" their clocks, the observers in *S'* will simply set a clock located at point *D* to mark

$$t_D' = L'/c \qquad (25)$$

when the light ray reaches *D*, because they have assumed the one-way light speed to be *c*. The internal "synchronization" scheme is shown in figure 4. It uses a procedure to relate the times of distant clocks in an inertial frame *as if* the moving frame was Einstein's frame.

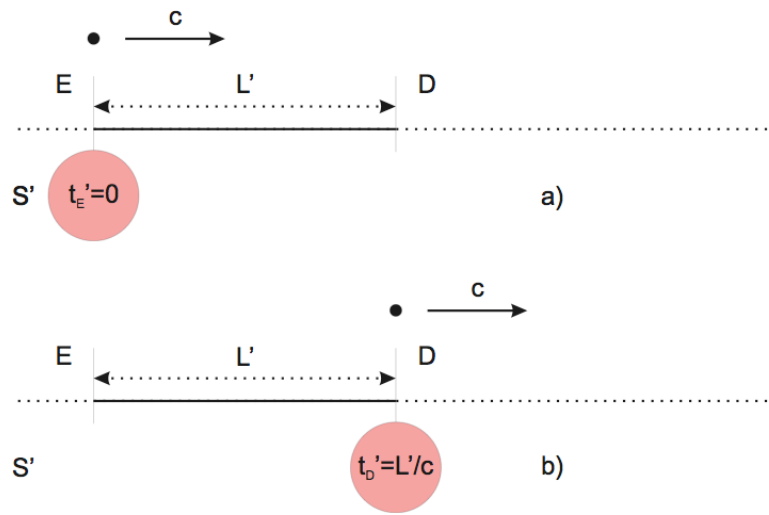

Figure 4: Internal "synchronization" of moving clocks: a light ray (represented by the black dot): *a)* is emitted from *E* at $t_E'=0$; and *b)* arrives at *D*, setting a clock placed there to mark $t_D'=L'/c$.

Do these clocks establish the common time for the moving frame? Evidently not: the time readings of the internally "synchronized" clocks can be treated as "times", but they are true times only *if* the one-way speed of light is indeed *c*. From the previous sections we know two important things: first, that different types of clocks provide different time coordinates to describe the same reality; and second, that the words "time", "speed" and "simultaneity", to which we use to attribute a precise physical meaning, actually refer to different notions when different types of clocks are used. Of course we have stressed several times that many different descriptions, made with various types of clocks and rulers, are mathematically equivalent[4]. Thus, this latter issue is mainly a question of language, although an important one and rather likely to originate severe misunderstandings. Because the physical concepts underling each of these descriptions are quite different. For these reasons, it is of major importance to know with what kind of clocks one ends up after performing an internal "synchronization".

---

[4] Notice, however, that it is not possible to transform one type of coordinates to another one, for instance from Lorentzian to synchronized coordinates, unless the absolute speeds of the moving frames are known. This question is addressed below.

It is not too difficult to realize that *internally "synchronized" clocks are de-synchronized Lorentzian clocks*. In the previous section the speeds measured with Lorentzian clocks were designated by Einstein speeds. It has been subsequently seen that the one-way Einstein speed of light is $c$ in all inertial frames (*cf.* equation (21)). As a consequence, when the internal "synchronization" is done and the one-way speed of light is imposed to be $c$, Einstein speeds and Lorentzian clocks are in fact being used! There is no problem with it, as long as we are aware we are doing so. Moreover, it may be even necessary to proceed in this way, in particular if it is not possible to look outside or to identify Einstein's frame. But it always must be kept in mind that Lorentzian times are not "true" times, Einstein speeds are not the "true" speeds, and internally "synchronized" clocks are not synchronized. Since it should be already clear what exactly an internal "synchronization" means, we can drop the "" signs from now on.

Let us check how the internal synchronization of the moving clocks in $S'$ is seen from Einstein's frame $S$. For simplicity, assume the light ray is emitted from $E$ at $t=0$, as shown in figure 5.

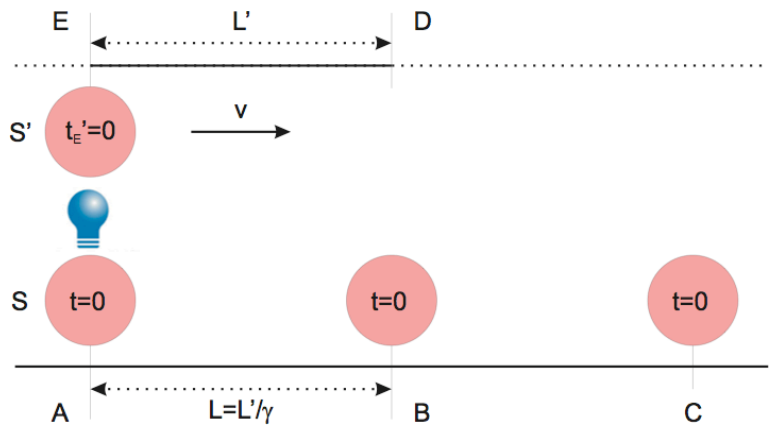

Figure 5: Internal synchronization of moving clocks: a light ray is emitted from $E$ at $t_E'=t=0$...

When does the light ray reach $D$, for the observers at rest? Since in $S$ the relative speed of light and the clock in $D$ is $c-v$, light takes $L/(c-v)=L'/[\gamma(c-v)]$. Hence, the clocks in $S$ mark

$$t_1 = L'/\gamma\,(c-v) \qquad (26)$$

when the light ray reaches $D$. What are the time readings of the moving clocks $E$ and $D$ at this time? Clock $D$ was prescribed to mark $t_D'=L'/c$, as given by (25). On the contrary, clock $E$ was running already since $t=0$, and is affected by time dilatation. It is then showing

$$t_E' = t_1' = t_1/\gamma \qquad (27)$$

Substituting $t_1$ and $\gamma$, this last expression can be written in the form

$$t_E' = (L'/\gamma^2) \times 1/(c-v) = L'(1-v^2/c^2)/(c-v) = (L'/c^2)(c^2-v^2)/(c-v) = (L'/c^2)(c+v) \quad (28)$$

Finally,

$$t_E' = L'/c + L'v/c^2 \quad (29)$$

The arrival of the light signal at $D$ is represented in figure 6.

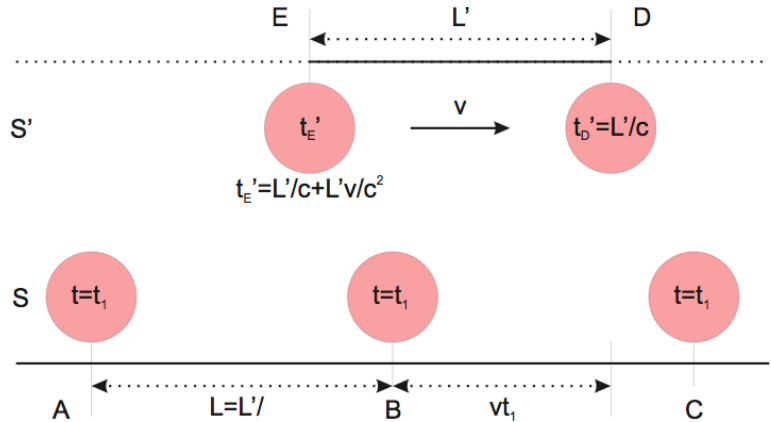

Figure 6: ... and arrives at $D$, setting a clock placed there to mark $t_D' = L'/c$.

Notice that if clock $D$ was synchronized, it should be marking the same time $t_1' = t_E' = L'/c + L'v/c^2$ as clock $E$. However, comparing (25) with (29), it is concluded that

$$t_D' = t_E' - L'v/c^2 \quad (30)$$

which confirms that *clock $D$ is de-synchronized precisely according to the definition of Lorentzian time* given by (15). In this way, the internal synchronization scheme assuming the one-way light speed to be $c$ in a moving inertial frame constitutes an alternative way to obtain Lorentzian times and the Lorentz transformation.

There is an easy analogy between the internal synchronization of clocks using light speed *as if* it was $c$ in all directions and "synchronization" of clocks around a race track using a $F_1$ car *as if* its speed was constant. Suppose a $F_1$ car is going in a circuit, doing a few laps exactly in the same way. Someone is standing with a clock on the start/finish line, and registers the time the $F_1$ takes to make one lap. Knowing the length of the track, it is easy to find the average speed of the $F_1$ during the lap. For instance, if the $F_1$ takes 1 minute and 12 seconds (*i.e.*, 0.02 hours) to complete a 5 km long circuit, it is racing at the average speed of 5/0.02=250 km/h. This time measurement is of course made with only one clock, located at the start/finish line. In respect to the light synchronization of clocks, this first measurement is equivalent to verifying that the two-way speed of light – the average speed of light in a round trip – is actually $c$. Next, imagine that several other observers are sitting in some other spots of the circuit. At a certain arranged lap, the person on the start/finish line sets his clock to mark zero when the $F_1$ crosses the line. Then, each of the other observers sets his own clock to mark "distance of his location to the start/finish line" over 250 km/h when the F1 passes. This corresponds to "synchronizing" the clocks *as if* the $F_1$ speed was constant and equal to 250 km/h all way around the circuit, even if it is not. For example, a person located 1 km after the start/finish line sets his clock to 1/250=0.004 h = 14.4 seconds. In the end all observers have their clocks "internally synchronized with the speed of the F1". Obviously the real speed of the $F_1$ is not 250 km/h in all parts of the circuit. But from now on, its "speed" measured with these clocks will always be 250 km/h. Because the clocks have been set using such a procedure that it cannot be otherwise. The time readings of these clocks can be called "times" and can be used to perform time measurements. But they are likely to induce their users to arrive at odd interpretations. The same situation occurs with the internal synchronization of clocks with the speed of light. All observers in a moving frame can synchronize their clocks with light signals *as if* the one-way speed of light was constant and equal to $c$ in all directions, even if it is not. In the end they will have their clocks internally synchronized. And from now on, the one-way "speed" of light measured with these clocks – the Einstein speed of light – will always be $c$. Because the clocks have been set using such a procedure that it cannot be otherwise. As with the Grand Prix spectators, the time readings of these clocks can be called "times" and can be used to perform time measurements. But they are likely to induce their users to arrive at odd interpretations.

We are now ready to have a close look at Einstein's theory of special relativity. This theory is built from two postulates, the principle of relativity and the constancy of the speed of light. Let us check what they say. In his 1905 article [Einstein1905], Einstein starts with the definitions of simultaneity, synchronization and time for the rest system, as it has been seen in section 2. Subsequently, he verifies what happens when two moving observers, each carrying his own clock,

> apply to the two clocks the [same] criterion for the synchronous rate of two clocks.

Therefore, in Einstein's theory of relativity the observers in moving inertial frames proceed *as if* they were at rest. In particular, and in order to "synchronize" their clocks, all inertial observers assume the one-way speed of light in empty space to be $c$, independently of the state of motion of the emitting body. This last sentence is actually Einstein's second postulate. As a result, Einstein synchronization of clocks is the internal synchronization detailed above, corresponding to Lorentzian clocks. Einstein time intervals are the time intervals given by Lorentzian clocks. Einstein simultaneity is the "simultaneity" exhibited by Lorentzian clocks. Einstein speeds are the "speeds" measured with Lorentzian clocks, *i.e.*, precisely what we have defined as... Einstein speeds. And Einstein transformation of coordinates between inertial frames is the Lorentz transformation. All Einstein's definitions are extremely precise, clear and full of physical content. However, as discussed in the previous section, the words "synchronization", "simultaneity", "time lapse" and "speed" do not refer to the notions we generally have in mind when using them. They must be used with caution, because in a sense they become false-friends and more than often originate misinterpretations. Einstein's own conviction (see, *e.g.*, [Shankland1973]) was that his notions of "synchronization", "simultaneity", "time lapse" and "speed" were the only true "synchronization", "simultaneity", "time" and "speed" for the description of physical phenomena. This is why the language used became messy and a critical issue. Einstein actually redefined all these words, and they were given a meaning they have *not*. Notice that Einstein's theory is valid and can be used. But when some statement is made it does not usually mean what it is assumed it does! It must be always kept in mind that Einstein "synchronization" is the internal synchronization, Einstein "simultaneity" is the simultaneity given by Lorentzian clocks, Einstein "times" are Lorentzian times and Einstein "speeds" are Einstein speeds. For instance, when it is said "the one-way speed of light is $c$ is all inertial frames" no one really understands how can it be (see the comments in the excellent textbook by David Morin [Morin2003]). But the sentence "the one-way Einstein speed of light is $c$ is all inertial frames" is a complete triviality.

One of the features of Einstein's relativity of great consequence is the *relativity of simultaneity*. As stated by Einstein himself [Einstein1905],

> we cannot ascribe *absolute* meaning to the concept of simultaneity; instead, two events that are simultaneous when observed from some particular coordinate system can no longer be considered simultaneous when observed from a system that is moving relative to that system.

Most of the "paradoxes" and difficulties in interpretation arising in the theory of relativity are in fact related to this relativity of simultaneity (see for instance, the very interesting discussion of Nelson, Rowland and Mallinckrodt [Nelson2003, Rowland2004, Mallinckrodt2004, Nelson2004]). Relativity of simultaneity was pointed out in the previous section with the help of figure 3. It is simply a consequence of considering, by definition, de-synchronized Lorentzian clocks as being "synchronized", Lorentzian times as "times" and the associated operational notion of simultaneity as "simultaneity".

Einstein's first postulate, his principle of relativity, is related to the properties of the Lorentz transformation. In the previous section we have seen that the Lorentz transformation is symmetrical, keeps the same form for any two moving inertial frames, and involves only the relative Einstein speed of the two frames. Hence, in Einstein's theory of relativity, where Lorentzian times are taken as "times" and Einstein speeds are taken as "speeds", *all inertial frames are "equivalent"*. No inertial reference frame is better than any other. This is how Einstein's principle of relativity is presented in many textbooks. The equivalence of all inertial frames postulated in Einstein's relativity means that the laws of physics *keep the same form* in all inertial frames, as it happens with Newton's second law (14) when Galileo transformation of coordinates is used. Historically, of great impact and with a decisive contribution to the success of Einstein's theory of relativity was the fact that, in contrast with the situation with Galileo transformation, Maxwell's equations of electrodynamics keep the same form when the Lorentz transformation of coordinates is used. This was actually Einstein's own formulation of the principle of relativity, as we have quoted in section 4. The impossibility of detecting absolute motion is then a *consequence* of this equivalence of all inertial frames. Because, if all frames are "equivalent", it makes no sense to say that something *is* moving: it only makes sense to say that one thing is moving with respect to another. In this way, in Einstein's relativity *the equivalence of all inertial frames is more fundamental than the impossibility of detecting absolute motion*. The idea of absolute rest and absolute motion become superfluous and can be abandoned.

Taking into account what we have presented so far, Einstein's principle of relativity can be reinterpreted under a new light. In section 4 we have formulated the principle of relativity as

> All the experiments performed in a closed cabin in vacuum in any moving inertial frame will appear the same as if performed in Einstein's frame, provided, of course, that one does not look outside.

We stressed the importance of not looking outside. As a matter of fact, *if* one does not look outside, the synchronization of clocks must be done internally. And internally synchronized clocks are Lorentzian clocks. Therefore, all moving inertial frames, when equipped with Lorentzian clocks, *appear* to be "equivalent". All experiments and all measurements made with Lorentzian clocks in a moving frame must give the same result as if they were made in Einstein's frame (which is the only frame where synchronized clocks and Lorentzian clocks are the same). Our principle of relativity can hence be rewritten in the following way:

> All laws of physics, when written with Lorentzian coordinates – *i.e.*, with Lorentzian times and Einstein speeds – keep the same form in all inertial frames, the same as in Einstein's frame.

This is the meaning of Poincaré's "nature conspiracy". With Lorentzian coordinates, any moving inertial frame appears to be Einstein's frame. In this way, *the impossibility of detecting absolute motion without looking outside is more fundamental than the supposed equivalence of all inertial frames*, as emphasized by authors like Fock [Fock1955]. This "equivalence" of all inertial frames is merely formal and no more than a result of using Lorentzian coordinates. The laws of physics only *keep the same form* when written with Lorentzian coordinates. If they are written with other coordinates, such as the synchronized coordinates, they do *not* keep the same form. But of course the laws of physics *are the same* in all inertial frames! Physics and its laws do not change with the coordinates we chose to describe reality.

It should be mentioned that Feynman [FSL1979] makes a not so traditional presentation of Einstein's special relativity and writes the principle of relativity correctly. As noted before, he is among the very few to have stressed the importance of not looking outside. Moreover, still more strikingly, he discusses the "invariance" of the laws of physics in the right way:

> *all the physical laws* should be of such a kind that they *remain unchanged under a Lorentz transformation*.

Contrary to most textbooks, Feynman does not simply state that the laws of physics keep the same form in all inertial frames, but specifically mentions the crucial role of the Lorentz transformation. The Lorentz transformation has a magic aura in physics due to its mathematical properties of invariance of the laws of physics. Which, as it has been seen, are strongly connected with the fact that the Lorentz transformation is the natural transformation of coordinates that arises when a moving inertial frame is treated *as if* it was the rest system.

There are evidently two very important issues. In one hand, to know if and how it is possible to look outside. On the other hand, to know what happens if one *does* look outside. These matters are discussed in [AG2006a] and are only briefly addressed below. Remarkably, the answer to the question "what is meant by looking outside" is twofold and far more subtle than it might be thought at first sight, but this discussion is not made here.

It is still worth mentioning that Einstein's relativity, with its equivalence of all inertial frames, is the theory of the "points of view". The real situations – and ultimately reality! – do not look like one thing in particular. As put by the classic book by David Morin [Morin2003],

> There is no such thing as "is-ness", since the look depends on the frame in which the looking is being done.

Of course things depend from the frame in which the looking is being done. As mentioned in the quote of Feynman in the beginning of last section, a person looks different from the front than from the back. But Morin's is-ness refers to much more unusual and relevant things. Within the framework of Einstein's relativity, where Lorentzian times and Einstein speeds are considered to be "true" times and speeds, it makes no sense to say things like "two events are simultaneous", "clock *A* runs slower than clock *B*" or "train *A* is longer than train *B*". The answer to these questions in Einstein's relativity depends upon one's point of view. But notice once more that the Lorentz transformation, with its lack of is-ness of reality, is mathematically equivalent to the synchronized transformation, with its absolute assertions about simultaneity, clock rhythms and length measurements. As long as the absolute speeds of the moving frames are known, it is immediate to transform Lorentzian times and Einstein speeds into true times and true speeds. However, if these absolute speeds are unknown, then it is impossible to convert the Lorentzian times and Einstein speeds resulting from an internal synchronization into true times and speeds...

Can it be determined which is Einstein's frame? One might be tempted to answer that it is enough to measure the one-way speed of light. Its value in vacuum is *c* in all directions only in Einstein's frame. This answer is correct... but useless. Because according to our understanding of the principle of relativity, it seems such a measurement cannot be made. As a matter of fact, the synchronization of the clocks in Einstein's frame is an *internal* one. And in vacuum, using only internal procedures, any inertial frame appears to be Einstein's frame! In this way, in vacuum there is no way to determine which is Einstein's frame. Which implies there is no way to make a real synchronization of clocks, to measure the one-way speed of light nor to determine an absolute speed. It is again Poincaré-Feynman nature conspiracy, preventing any observer to determine an absolute speed. Is it really impossible to make such determination? The question is still open, and this is may be not the case, as indicated by the works of Cahill [CK2002, Cahill2004] and Consoli [CC2003, CC2004] and discussed in [AG2006a]. Although in vacuum it is, in harmony with the principle of relativity.

The (eventual) impossibility of detecting experimentally the "rest system" does not change anything in Lorentz-Poincaré philosophy. Within its framework, one would simply have to admit he does not know the absolute speed of an object, but this does not promote the Einstein speed to the status of ("true") speed. They remain different notions. Consequently, the scenario of a preferred frame keeps being consistent even if this frame cannot be identified. If Poincaré's nature conspiracy holds, then it is impossible to experimentally determine the one-way speed of light. And *without the measurement of the one-way speed of light relativity theories remain undetermined and incomplete*, as previously noted in [Abreu2004]. All it can be done is to proceed "as if". For instance, one option is to proclaim which is the inertial frame *S'* to be treated *as if* it was Einstein's frame. And "synchronize" its clocks *as if* the one-way speed of light in it was *c*. Then, if it is possible to look outside by actually looking through a window to what is outside, the clocks from the other inertial frames can be "externally synchronized" with the help of the clocks from *S'*. The relationships between the physical quantities in the different frames are then the same derived in section 2, with *S'* playing the role of Einstein's frame *S*. However, even if these relationships are the same, the meaning of the measured physical quantities is not! As shown in the previous sections, one can perform measurements with clocks "synchronized" considering *S'* as if it was Einstein's frame. But, not knowing the absolute speed of *S'*, it is not possible to translate the Lorentzian coordinates into synchronized ones. In this case, the notions of "time", "speed" and "simultaneity" can only be given the meaning of Lorentzian time, Einstein speed, and simultaneity measured with Lorentzian clocks. Similarly to the $F_1$ car synchronization example from the previous section, in which it is impossible to infer the real speed of the $F_1$ car in any part of the circuit from a "speed" measurement, without the knowledge of which is Einstein's frame it is not possible to know the real speeds and time intervals from the "speeds" and "time intervals" actually measured with Lorentzian clocks.

Another option is not to look outside at all and to treat *all* moving inertial frames *as if* each of them was the rest Einstein's frame. From the previous section, this is evidently Einstein's relativity. In this case all moving frames become furnished with Lorentzian clocks and thus measure Einstein speeds. Once more, it is necessary to know the absolute speeds of each inertial frame in order to be able to translate the Lorentzian coordinates into synchronized ones. It is necessary to know which is Einstein's frame if we want to ascribe an absolute significance to the quantities that are being measured. But this is precisely what we cannot know. And so relativity theories are indeed undetermined and incomplete. It is possible to proclaim one inertial frame to be treated *as if* it was the rest frame. It is even possible to proclaim that all inertial frames will be treated *as if* they were the rest frame. It is possible to do measurements under these assumptions. But the measured Lorentzian times and Einstein speeds cannot be translated into real times and real (or "absolute") speeds. Einstein's relativity gives us an *operative procedure* to study *relative* motion without knowing the absolute speeds nor the actual one-way speeds of light. But this is fully compatible with and can be understood from the theory presented here, a theory of *absolute* motion that puts into perspective the assertions made within special relativity.

## 6. An example: time dilation

Time dilation was illustrated first in section 2 with the help of figures 1 and 2. It was subsequently presented in section 3 and figure 4, for the case of a moving frame furnished with Lorentzian clocks. These three figures are merged here in figures 6 and 7, which show a moving inertial frame *S'* equipped both with synchronized and Lorentzian clocks. As it has been seen in the previous sections, time dilatation is a question of rhythms of clocks, independent from the type of "synchronization" that is used. In figures 6 and 7, 8 milliseconds pass for each of the moving clocks *D* and *E*, while 10 ms have passed for each of the clocks at rest *A*, *B* and *C*. Moving clocks run slower. In this case, by a factor of 10/8=1.25. The affirmation does not depend on the initial adjustment that is made to the moving clocks, *i.e.*, on the type of clocks that is being used.

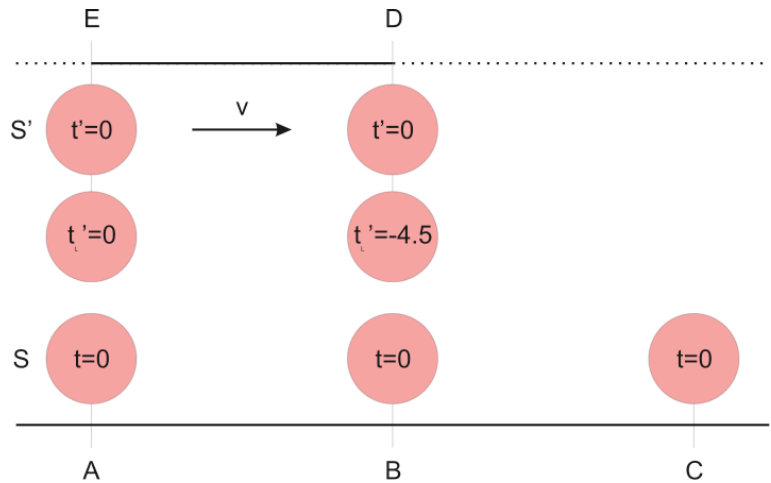

Figure 6: Time dilatation both with synchronized and de-synchronized Lorentzian clocks. All times in the figures are expressed in ms. The initial situation at *t*=0...

Let us now see how the observers from *S'* describe the situation depicted in figures 6 and 7. Well, if they use their synchronized clocks, or if they use their Lorentzian clocks and they know the speed of *S'* in order to translate the Lorentzian coordinates into synchronized ones, they simply say the same as the observers from *S*: 8 milliseconds passed in *S'* while 10 milliseconds elapsed in *S*. There is nothing more about it.

Clearly, there is no reciprocity between frames: since motion is an absolute notion, the clocks from Einstein's frame are truly at rest, even if they can be seen as moving in relation to *S'*. But time dilatation is a consequence of (absolute) motion, not of *relative* motion. Therefore, clocks in *S* are not affected by time dilatation.

However, if the observers in *S'* do not know they are moving, if they do not look outside hence using their Lorentzian clocks, they can proceed *as if* they were at rest. They would then consider themselves "at rest" and Einstein's frame *S* to be "moving". An observer with clock *D* would say he had seen a clock *B* from *S* just in front of him, and that this clock *B* was reading *t*=0 while his clock *D* was marking $t_L'$ = -4.5 ms. Later on, an observer co-punctual with clock *E* would see the same clock *B* showing *t*=10 ms, while his own clock *E* would exhibit $t_L'$ = 8 ms. The observers from *S'* would then (erroneously) conclude that while for them, "at rest", 8-(-4.5)=12.5 ms have passed, in the "moving" frame *S* only 10-0=10 ms have passed. Therefore, the "moving" clocks from *S* would appear to run slower! In this case, by a factor 12.5/10=1.25, as it had to be. However, *this effect has nothing to do with the rhythms of clocks* and it is merely a result of comparing time coordinates using de-synchronized Lorentzian clocks. This example shows well how dangerous it is to use de-synchronized Lorentzian clocks while convinced they are truly synchronized. A clock running slower can even be thought to be running faster. But there is no contradiction between both descriptions: they are both valid, it is "only" that their statements – which indeed seem contradictory – refer to different notions.

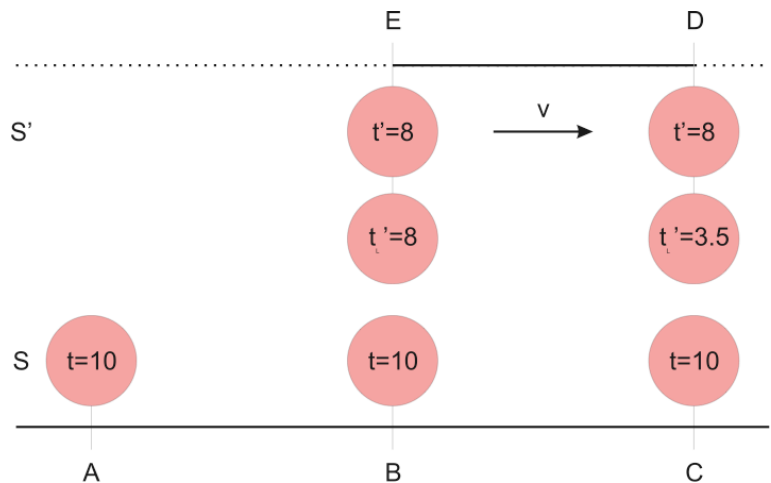

Figure 7: ... evolves and it is shown at *t*=10 ms.

The question of "reciprocity" or "symmetry" between frames is very interesting. As a matter of fact, it has been previously mentioned that, when Lorentzian coordinates are used, all inertial frames *appear* to be "equivalent". Einstein's relativity considers all inertial frames to be "equivalent", so that the observers from *S'* have to see *S* in the same way as the observers from *S* see *S'*. This is true only if *S* and *S'* use Lorentzian clocks, Lorentzian time intervals and Einstein speeds. With lorentzian time coordinates, even time dilation appears to be symmetric and a *relative* effect. Notice that *one* clock from *S* apparently gets delayed in relation to the *several* Lorentzian clocks from *S'* it crosses. The observers from *S'*, considering themselves "at rest", say the "moving" clock *B* is getting delayed. And they can use a similar argument for *each* clock from *S* individually, but not for the clocks of *S* as an all. Similarly, *one* clock from *S'* apparently (and truly) gets delayed in relation to the *several* clocks from *S* it crosses. The observers from *S*, which are at rest, say, for instance, the moving clock *E* is getting delayed. And they can use a similar argument for *each* clock from *S'*. Thus, the *description* of time dilation that is made with Lorentzian clocks is indeed symmetrical between *S* and *S'*. But the *phenomenon* of time dilation is not. Finally, with Lorentzian clocks, this analysis is the same for any two moving inertial frames *S'* and *S"*. And, clearly, without the knowledge of which is Einstein's frame, without the ability to "look outside", when two inertial frames cross it is possible to compare their time coordinates, but it is impossible to know in which frame clocks are actually running slower (!). Evidently all this can be rather confusing, especially if Lorentzian times are seen as "times" and if the mere comparisons of time coordinates are taken as the "rhythms" of clocks.

## 7. Conclusions

The compatibility between Special Relativity and Lorentz-Poincaré view of a preferred reference system experimentally inaccessible was shown to hold and exemplified with the case of time dilation. The consistency of both scenarios is thoroughly discussed in [AG2006a], where many other classic examples are illustrated, such as the "reciprocity" of space contraction, the twin paradox, the problem of Bell's accelerating spaceships, the propagation of spherical electromagnetic waves and the electric field of a moving point charge. The later examples show that the present theory, although presented in a quite simple and accessible way based on very fundamental examples, is not restricted to kinematics. On the contrary, its application to dynamics and electromagnetism is straightforward.

The key point is that reality can be described in many different ways, for instance using "synchronized", "Galilean" (see [AG2006a, AG2006b]) or "Lorentzian" clocks, which are all mathematically equivalent. However, a change in the *description* does not change reality itself. Nevertheless, there may exist difficulties in assigning a physical meaning to the quantities measured with clocks other than the synchronized ones, since some words to which we are used to ascribe a precise notion, such as "time interval", "simultaneous", "synchronous" and "speed", correspond to different things when measured with different types of clocks. Any of these notions can be redefined, but one must then be aware they do not coincide anymore to what usually it is thought they do. The "contradiction in terms" between Special Relativity and Absolute Space is no more than a terminological confusion.

"Speeds" measured with de-synchronized Lorentzian clocks are denoted as Einstein speeds. The one-way Einstein speed of light is always $c$ in a moving inertial frame, regardless of the (absolute) speed of the moving frame. Lorentzian clocks are thus obtained when clock "synchronization" is made internally, i.e., when the moving frame is treated as if it was Einstein's frame and the one-way speed of light is assumed to be $c$ for operational purposes, even if it is not. Notice that the question of the "constancy" of the one-way "speed" of light is essentially an issue of *language*, related to which description is being used. Therefore, when presenting or discussing Special Relativity we must keep in mind the precise meaning in which words like "speed" and "simultaneity" are used. Notice that this remark goes deeper than the *conventionalist* thesis and a vision of physics based on *operationalism*.

The generalized adoption of a careless and incorrect language, pretending what can be affirmed is something that cannot be, is a critical issue. For instance, the one-way speed of light in a particular moving inertial frame is not changed because the one-way Einstein speed of light is $c$. Claiming that the difficulty (or even impossibility) of knowing the one-way speed of light in a particular inertial frame can be solved by stating the one-way Einstein speed of light is $c$ in all inertial frames is of little signification.

The principle of relativity – or principle of relative movement – is the statement of the impossibility of detecting absolute motion in vacuum without looking outside: all experiments performed in a closed cabin in vacuum in any moving inertial frame will appear the same as if performed in Einstein's frame, provided that one does not look outside. With the Lorentzian clocks resulting form an internal synchronization, any inertial frame appears to be Einstein's frame. Since the synchronization of the clocks in Einstein's frame is also an internal one, the principle of relativity implies the impossibility of detection of Einstein's frame in vacuum and the consequent indeterminacy of relativity theories.

Einstein theory of relativity results from "synchronizing" the clocks of each moving inertial frame as if that frame was Einstein's frame, *i.e.*, to perform an internal synchronization and to use Lorentzian clocks in all moving inertial frames. The principle of relativity then means that all laws of physics keep the same form when written with Lorentzian coordinates. All inertial frames appear to be "equivalent" when Lorentzian coordinates are used. Nevertheless, neither the principle of relativity nor the apparent equivalence of all inertial frames are incompatible with the reality of a preferred, and thus absolute, frame.

For the view presented here, it is not crucial if Einstein's frame can actually be detected experimentally. Nevertheless, it seems the identification of Einstein's frame may be possible by releasing the constraint of being in vacuum, as a consequence of an interaction between moving media and the rest Einstein's frame. Such an interaction, if confirmed, provides a subtle and elaborated way of "looking outside". The non-null result of Michelson-Morley experiments suggests such an interaction does exist and the identification of Einstein's frame can be done [CK2002, CC2003, Cahill2004, CC2004]. If this is the case, the principle of relativity can be generalized to moving media only as an approximation, being exact only in vacuum [AG2006a].

Besides the works by Franco Selleri [Selleri1996, Selleri2005], there is another very nice article with a message relatively similar to the one conveyed here, published by Leubner and co-workers in 1992 [LAK1992]. After developing a non-standard synchronization, which they named "everyday synchronization" (and actually corresponds to the synchronized transformation for the particular case $v=-c$), they conclude:

> After these educational benefits of studying the set of standard 'relativistic effects' also in everyday coordinates, we are of course happy to drop again 'everyday' synchronization before proceeding to less elementary aspects of relativistic physics. For these aspects, we certainly prefer the more symmetric coordinate representations of expressions resulting from Einstein synchronization, but on purely practical grounds, and not on philosophical ones.

We subscribe both Bell's and Leubner's ideas, and hope in the near future physics textbooks will at least regularly include an analysis of Lorentz's philosophy and of its consistency, assuming without any prejudice that "the facts of physics do not oblige us to accept one philosophy rather than the other". But we aim more. What we have shown is that there is no conflict between the Lorentz-Poincaré and the Einstein-Minkowski programmes; there is no need to "chose" between one and the other: the latter is in fact a particular aspect of the former, obtained by an operative procedure that allows the study of relative motion without the knowledge of Einstein's frame.